\documentclass[preprint,aps]{revtex4}

\usepackage{amsmath,amssymb,amsthm,graphicx,latexsym}

\newcommand{\be}{\begin{equation}}
\newcommand{\ee}{\end{equation}}
\newcommand{\bea}{\begin{eqnarray}}
\newcommand{\eea}{\end{eqnarray}}
\newcommand{\bml}{\begin{subequations}}
\newcommand{\eml}{\end{subequations}}
\newcommand{\bfig}{\begin{figure}}
\newcommand{\efig}{\end{figure}}

\newcommand{\bg}{\beta}

\newcommand{\ve}{\varepsilon}

\newcommand{\thg}{\theta}

\newcommand{\lb}{\lambda}
\newcommand{\sg}{\sigma}

\newcommand{\vf}{\varphi}

\newcommand{\Gam}{\Gamma}

\newcommand{\Og}{\Omega}
\newcommand{\Lb}{\Lambda}

\newcommand{\lh}{\left(}
\newcommand{\rh}{\right)}

\begin{document}

\pagestyle{empty}

\begin{center}
{\bf {\Large Generalized particle dynamics in anti de Sitter spaces:
A source for dark energy}}\\
\vspace{2ex}

Sudipta Das$^{a}$,
Subir Ghosh$^{a}$,
Jan-Willem van Holten$^{b}$
and Supratik Pal$^{a, c}$
\vspace{2ex}

$^a$Physics and Applied Mathematics Unit, Indian
Statistical Institute, \\
203 B.T.Road, Kolkata 700108, India

$^b$NIKHEF, PO Box 41882, 1009 DB Amsterdam, Netherlands

$^c$Bethe Center for Theoretical Physics and Physikalisches Institut der Universit\"{a}t Bonn, Nussallee 12, 53115 Bonn, Germany
\end{center}

\begin{abstract}
We consider the generalized particle dynamics, proposed by us
\cite{gpdsch}, in brane world formalisms for an asymptotically
anti de Sitter background. The present framework results in a
new model that accounts for the late acceleration of the universe.
An effective Dark Energy equation of state, exhibiting a phantom
like behavior, is generated. The model is derived by embedding
the physical FLRW universe in a $(4+1)$-dimensional effective
space-time, induced by the generalized particle dynamics. We
corroborate our results with present day observed cosmological
parameters.
\end{abstract}

\maketitle

\section{\bf introduction} \noindent

In recent times theoretical understanding of Dark Energy (DE) has
become the Holy Grail of cosmological investigations. Existence of
DE is unavoidable if one wants to explain the (present day)
accelerated expansion of the universe. However, if the dynamical
laws of motion pertaining to General Relativity are held sacred,
it seems inevitable that a paradigm shift in the properties of DE
constituent is needed. This is simply because normal matter creates
a positive pressure that decelerates the universe expansion
whereas DE has to generate a negative pressure to
enhance the expansion.

Observational vindication \cite{snsnls} of
the present-day acceleration of the universe, and the subsequent
precise measurements of observable parameters \cite{hstcmb, wmap5,
shoes} indicate that the entity Dark Energy (DE) which is
responsible for the recent acceleration contribute to $\sim 70 \%$
of cosmic energy density. This DE density-fraction of total cosmic
density ($\Omega_{\rm DE}$) is determined conclusively from
several independent probes ($\Omega_{\rm DE} = 0.726 \pm 0.015$ at
$95 \%$ C.L. from latest WMAP5 data \cite{wmap5}). On the other
hand, DE {\em effective} Equation of State (EOS), $w_{\rm DE}$, is still inconclusive.
But the big surprise is that this agent seems even more exotic in
nature than imagined before due to the fact that most likely its
EOS crosses the so-called Phantom Divider
(i.e. $w_{\rm DE} = -1$). Though SNLS data show no general
behavior for $w_{\rm DE} < -1$, the analysis of the most reliable
SNIa Gold dataset show strong indication that $w_{\rm DE} < -1$
\cite{pdlobs} (the lower bound being $-1.11 < w_{\rm DE}$ from
WMAP5 data \cite{wmap5}), leading to the conclusion that models
with phantom divider crossing are preferred over $\Lambda$CDM (or
quintessential candidates) at $2 \sigma$ level. This clearly
weakens the claims of cosmological constant $\Lambda$ or dynamical
models like quintessence, Kessence, Chaplygin gas etc. \cite{de}
as viable DE models. One can resurrect the scalar field models
only at the cost of phantom fields (quintom models) \cite{pdlth},
with a negative kinetic energy term but they bring in severe
instability problems and are better avoided.

In this perspective, instead of looking for contrived and
phenomenologically motivated DE models, it seems reasonable to
explore modified gravity theories \cite{pdlmod, brde} which do not
suffer from any such major drawbacks. But one might still feel
skeptical since more often than not explicit forms of these
modified gravity theories appear to be designer made and/or fine
tuned without a deeper dynamical framework based on first
principles. In a previous work \cite{gpdsch}, we have already constructed a lower
dimensional toy model cosmology where the accelerated expansion
emerges in a brane world scenario. The modified gravity provides
the metric of the higher dimensional spacetime in which the brane
is embedded. (This well established formalism \cite{bwg} is explained below
and later as we proceed.) The interesting point is that the
specific modified gravity metric proposed by us can be derived in two ways: from
a generalized particle dynamics \cite{gpdsch} or from the
Kaluza-Klein type of reduction from a higher dimensional particle
model. We emphasize that in both these schemes conventional
relativistic dynamical principles are maintained and in the former
an extended form of spin-orbit coupling is introduced. In a
nutshell our toy model is an example of a successful union between
generalized particle dynamics and brane world frameworks. The only
limitation of our previous work \cite{gpdsch} was that it could not reproduce a
phantom like behavior simply because the bulk (background)
spacetime was asymptotically flat. This is rectified in the
present work as we describe below.

In this sequel we complete the project started in \cite{gpdsch} by
providing a variant of the Brane-world models where the novel
particle dynamics approach \cite{gpdsch} is extended to physical
3+1-dimensions. {\it{We demonstrate that the phantom-like behavior can
be induced without explicitly invoking the phantom field with
negative kinetic energy term.}} It is well-known (for details see
\cite{brde, gpdsch, bwg}) that by embedding techniques one can
relate cosmological surface dynamics (Friedmann equations) in
lower (e.g., 3+1) dimensions with particle motion in a higher
(e.g., 4+1) dimensional black-hole space-time. In present case, the latter
is taken as asymptotically anti de Sitter (AdS) space-time, taking
Schwarzschild-anti de Sitter (Sch-AdS) as a representative
example. In the standard brane world scenario the AdS background
induces an effective cosmological constant (which may be fine
tuned \cite{rs} or not \cite{brde, bwg}). However, a more
interesting situation occurs in our framework. Here the AdS bulk
induces a {\it{dynamical}} quantity which is essential in
imparting the phantom behavior. (This will become clear as we
proceed.) Indeed, apart from this bonus, there are strong
motivations for considering an AdS background: bulk spacetime in
RS model \cite{rs} is a slice of an AdS spacetime, the celebrated
AdS-CFT correspondence \cite{adscft} etc.. We demonstrate that in
the subsequent cosmological scenario, the induced effective
negative pressure can result in an expanding universe, capable of
crossing the phantom divide. Most notably, we further establish
our model by an analysis of the equation of state and a
determination of the relevant parameters describing the evolution
of the observable universe.

The paper is organized as follows: in Section II we introduce the
generalized particle dynamics in AdS space. Section III is devoted
to the Kaluza-Klein interpretation of the above. As cosmological
implications, we construct a Dark Energy model and a detail
quantitative discussion of this Dark Energy model is given in Section
IV. Section V consists of summary and future prospects of the
present work.

\section{\bf Generalized particle dynamics for AdS space}

Motivated by our previous work \cite{gpdsch}, in this paper we
intend to formulate a somewhat modified version of generalized
particle dynamics with non-minimal coupling for a particle moving
in ($4+1$)-dimensional asymptotically anti de Sitter (AdS) space-time,
taking Schwarzschild-anti de Sitter (Sch-AdS) as a representative
example. In the next section we will provide an
interpretation for this kind of particle dynamics by a
Kaluza-Klein decomposition. Further, we apply this to a
physical scenario in the cosmological context by embedding a
($3+1$)-dimensional FLRW universe into this ($4+1$)-dimensional
space-time. In the present section, however, we shall concentrate
on formulating the dynamics in the background asymptotically anti
de Sitter space.

Let us start with the
reparametrization-invariant action \cite{gpdsch}
\begin{equation}
S = \int L d\tau = m\, \int d\tau \left[ \frac{1}{2e} g_{\mu\nu}
\dot{x}^{\mu} \dot{x}^{\nu} - \frac{e}{2} - \lambda g_{\mu\nu}
\xi^{\mu} \dot{x}^{\nu} + \frac{e \lambda^{2}}{2} g_{\mu\nu}
\xi^{\mu} \xi^{\nu} +  \frac{e \beta \lambda^{2}}{2}\right].
\label{action}
\end{equation}
where $\tau$ is the worldline parameter, $\lambda(\tau)$ is an
auxiliary scalar variable, $e$ is the worldline einbein, $\beta$
is a numerical constant. Furthermore we demand the action to be
invariant under general coordinate transformation $\delta
x^\mu=\alpha\xi^\mu$, where $\xi^\mu$ are shown to be the Killing
vectors related to the symmetry of the spacetime \cite{holt}.
Clearly the first two terms in the action (\ref{action})
constitute the conventional particle action and the rest of terms
are introduced by us (see \cite{gpdsch}).

As already mentioned, our primary intention is to formulate
the dynamics for a particle moving in a ($4+1$)-dimensional
Sch-AdS space-time, the metric for which is given by
\be
g_{\mu \nu}dx^\mu dx^\nu =
-\left(k-\frac{2M}{r^2}+\Lb_5 r^2 \right)dt^2 +
\frac{dr^2}{k-\frac{2M}{r^2}+\Lb_5 r^2} + r^2 d\Omega_3^2,
\label{metric}
\ee
where $d\Omega_3^2$ is the three-sphere, $k ~( = 0,
\pm 1)$ is the curvature scalar and $\Lambda_5$ is the constant
curvature of the space-time. The action (\ref{action}) in this
($4+1$)-dimensional Sch-AdS space-time takes the form
\be S = \int L d\tau = \frac{m}{2}\, \int d\tau\, e \left[
\frac{1}{e^2}
 \lh - f(r) \dot{t}^2 + \frac{\dot{r}^2}{f(r)} + r^2 \dot{\vf}^2 \rh
 - \frac{2\lb}{e}\, r^2 \dot{\vf} + \lb^2 (r^2 + \bg) - 1 \right],
\label{i.7} \ee
where $f(r)=k-\frac{2M}{r^2}+\Lambda_5 r^2$ 
and $\xi^{\mu}=(0,0,0,0,1)$ is the Killing vector associated with the 
rotational symmetry of the metric. We have
restricted the motion of the particle on the equatorial plane, following usual practice.

The Killing vectors associated to this space-time lead to the
conservation of energy and angular momentum of a test particle of
mass $m$. Written in a convenient notation, this implies \be p_t
= - m f(r) \dot{t} = m \ve~~,~~ p_{\vf} = m r^2
(\dot{\vf}-\lambda) = m\beta \lambda=ml, \label{i.8} \ee where the
overdot represents a proper-time derivative. We have obtained
these relations by solving the equations of
motion for the worldline variables $e$ and $\lb$, whilst fixing
the gauge $e = 1$.

Along with (\ref{i.8}), we have an additional, modified mass-shell
constraint \cite{gpdsch} \be p^2+m^2+\frac{1}{\beta}(\xi.p)^2=0 \label{ms}
\ee These two equations are the key equations in governing our
formalism.

Using the expressions for momenta in this mass-shell constraint
the radial equation can be expressed in a convenient form \be
\dot{r}^2 + V_{\rm eff}(r) = \ve^2 \label{i.4} \ee where $V_{\rm
eff}$ is the effective potential, which, for the particle action
(\ref{i.7}), takes the form \be V_{\rm eff} = \lh k -
\frac{2M}{r^2} + \Lb_5 r^2 \rh \lh \mu^2 + \frac{l^2}{r^2} \rh.
\label{i.6} \ee by introducing a dimensionless parameter \be
\mu^2 = 1 + \frac{l^2}{\bg}. \ee We shall come to the implications
of this parameter shortly.

\begin{figure}[htb]
{\centerline{\includegraphics[width=9cm, height=6cm] {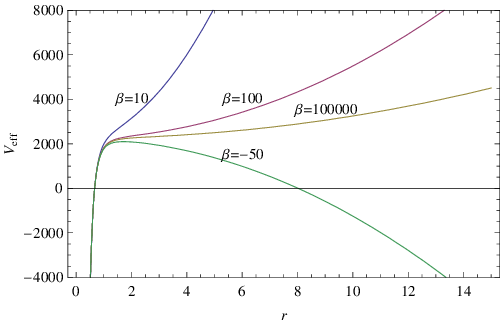}}}
\caption{Variation of the effective potential with radial distance
for different $\beta$ and $l=15$}
\label{figVr}
\end{figure}

In Figure \ref{figVr} we have plotted the effective potential
$V_{\rm eff}(r)$ against the radial coordinate $r$ choosing
different values for the parameter $\bg$. One can readily notice
from the plots that the effective potential is positive definite
for all positive values of $\bg$ whereas for negative values of
$\bg$ (with $l^2 > |\beta|$) the potential has a sign-flip, as it goes
again below the demarkation line zero.
As in \cite{gpdsch} we can readily interpret this distinctive
feature of the effective potential for negative $\bg$  as the
outcome of a {\it{repulsive force}} acting on the particle. This will
lead to interesting cosmological consequences, which will be
revealed in due course. This is the main interest of the present work.

The new parameter $\mu$ has an interesting physical significance.
To realize this, let us notice that
the effective potential can be re-written as
\be
V_{\rm eff} = F(r) \lh 1 + \frac{l^2}{r^2} \rh
\label{i.5}
\ee
with the induced radial function $F(r)$ defined as
\be
 F(r) = \lh k -
\frac{2M}{r^2} + \Lb_5 r^2 \rh \lh \frac{\mu^2 r^2 +l^2}{r^2 +
l^2} \rh. \label{i.2}
\ee
Incorporating the above redefinitions into the radial dynamics (\ref{i.4})
leads to a situation analogous to a pure Schwarzschild-AdS space-time
where the mass $m$ of a particle is modified as $m^2 \rightarrow \mu^2 m^2$.
This modification of mass can be realized by observing that
the mass-shell condition (\ref{ms}) can be written
now in the form $p^2 + m^2 \mu^2 = 0$.
Thus, one can say that a generalized particle dynamics
in a Schwarzschild-AdS background
is identical to a standard particle dynamics
in the background of an asymptotically AdS
space-time (with $m^2 \rightarrow \mu^2 m^2$)
where the induced metric is of the form
\be
ds^2_{4+1} = - F(r) dt^2 + \frac{dr^2}{F(r)} + r^2 d\Og_3^2
\label{i.1}
\ee
considering $c=1$. For $\mu^2 = 1$, equation (\ref{i.1})
reduces to the standard Schwarzschild-AdS metric.
We shall make use of the effective metric (\ref{i.1}) in deriving
the possible cosmological consequences later on in this article.

\section{\bf Kaluza-Klein interpretation}

The non-minimal particle dynamics
(\ref{i.7}) referred to in the previous section can indeed be
derived from a {\em minimal} model in a
space-time with one more spatial dimension, by imposing a
constraint on the additional momentum component $p_{d+1} = 0$ and
then performing Kaluza-Klein type decomposition. In this section
we shall derive this result in details for arbitrary spacetime dimensions.

Let us consider a particle of mass $m$ performing geodesic motion in a
($d+1$)-dimensional space-time with metric $G_{AB}$ ($A, B = 0,1,
...,d)$. The dynamics is described by the action
\be S = m \int
d\tau \lh \frac{1}{2e}\, G_{AB}\, \dot{x}^A \dot{x}^B -
\frac{e}{2} \rh
 \simeq - m \int d\tau \sqrt{- G_{AB}\, \dot{x}^A \dot{x}^B},
\label{1.1}
\ee
where the second expression follows by eliminating
$e$ by its equation of motion
\be
e^2 = - G_{AB}\, \dot{x}^A
\dot{x}^B. \label{1.2}
 \ee
The canonical momenta corresponding to the above action are given by
\be
p_A = \frac{m}{e}\, G_{AB}\, \dot{x}^B \simeq
 \frac{m\, G_{AB}\, \dot{x}^B}{\sqrt{-G_{MN} \dot{x}^M \dot{x}^N}}.
\label{1.3}
\ee
They satisfy the Hamiltonian constraint \be
\frac{2m}{e}\, H = G^{AB} p_A p_B + m^2 = 0. \label{1.4}
\ee
The Hamiltonian equations of motion can be written as
 \be \dot{x}^A =
\left\{ x^A, H \right\}~~~,~~~\dot{p}_A = \left\{ p_A, H \right\},
\label{1.5}
\ee
which readily gives the covariant equation
\be \dot{p}_A -
\dot{x}^B\, \Gam_{BA}^{\;\;\;C}\, p_C = 0. \label{1.6}
\ee

We know that if the
space-time geometry possesses isometries, there exist Killing
vectors $\xi^A$ which satisfy the relation
\be
D_A \xi_B + D_B \xi_A = 0.
\label{1.7}
\ee
Then one can construct constants of motion of the following form
\be
C[\xi] = \xi^A p_A~~~,~~~\Rightarrow~~\dot{C}[\xi] = 0.
\label{1.8}
\ee

Employing Kaluza-Klein decomposition, let us now consider a special
form for the space-time with metric
\begin{equation}
G_{AB}=\left(
\begin{array}{cc}
  g_{\mu\nu} & - n_{\mu} \\
  - n_{\nu} & n^2 + \bg \\
\end{array}
\right)~~~,~~~(\mu,\nu)=(0,1,...,d-1) \label{2.1}
\end{equation}
where $g_{\mu\nu}$ is the metric for the $d$-dimensional subspace,
$n_{\mu} = g_{\mu\nu} n^{\nu}$  the normal vectors, and $n^2 =
g_{\mu\nu} n^{\mu} n^{\nu}$. Also, $\bg$ is just a numerical
constant which takes care of the consistency of the theory. Then
the inverse metric takes the form \be G^{AB} =\left(
\begin{array}{cc}
  g^{\mu\nu}+ \frac{1}{\bg}n^{\mu} n^{\nu}  & \frac{1}{\bg}\, n^{\mu} \\
  \frac{1}{\bg}\, n^{\nu}  & \frac{1}{\bg} \\
\end{array} \right) \label{2.4} \ee
where $g^{\mu\nu}$ is the usual $d$-dimensional inverse of
$g_{\mu\nu}$. We further specialize to the case where all
components are independent of $x^d$, implying
 \be G_{AB,d} = 0~~
\Rightarrow ~~ g_{\mu\nu,d} = 0~~,~~ n_{\mu,d} = 0. \label{2.2}
\ee
Consequently, $g_{\mu\nu}$ depends only on the co-ordinates $x^{\mu}$
and is the universal metric on all $d$-dimensional subspaces $x^d
=$ constant. Technically speaking,
 \be
 dx^d = 0 ~~~ \Rightarrow ~~~ G_{AB} dx^A dx^B =
g_{\mu\nu} dx^{\mu} dx^{\nu}. \label{2.3}
\ee
It follows that
there is translation invariance in the $x^d$ direction, generated
by the Killing vector $\xi^A_{(d)} = (0, ..., 0, 1)$, and the
momentum component in the $x^d$-direction is conserved.
Hence
 \be
\dot{p}_d = 0 ~~~ \Rightarrow ~~~ \xi^A_{(d)}\, p_A = p_d =
\mbox{constant}. \label{2.5}
\ee
which can be shown explicitly
using the equation of motion.
 Finally, we assume the
$d$-dimensional subspace to have an internal isometry generated by
a $d$-dimensional Killing vector $\xi^{\mu}$. Then we can specify
the off-diagonal metric components to be given by the covariant
components of this Killing vector: $n_{\mu} = \xi_{\mu}$, which
implies \be n_{\mu;\nu} + n_{\nu;\mu} = \xi_{\mu;\nu} +
\xi_{\nu;\mu} = 0, \label{2.9} \ee where the semicolon denotes a
$d$-dimensional covariant derivative. Defining the relation \be
\frac{1}{e}\, \dot{x}^d = \lb \label{2.10} \ee the action
(\ref{1.1}) reduces to \be S = m \int d\tau \left[ \frac{1}{2e}\,
g_{\mu\nu}\, \dot{x}^{\mu} \dot{x}^{\nu}
 - \frac{e}{2} - \lb \xi \cdot \dot{x} + \frac{e}{2} \lb^2 \lh  \xi^2 + \bg \rh
 \right],
\label{2.11} \ee
which is precisely the action of generalized
particle dynamics proposed in Eq (\ref{action}). Note that taking $\lb$
as a fundamental variable rather than $x^d$ does change the
Euler-Lagrange equations of motion, as we loose a time-derivative.
The net effect is, however, to constrain the $d$-component of the momentum
to vanish: \be p_d = 0, \label{2.12} \ee which is a particular
solution of eq.\ (\ref{2.5}). Indeed, one can conclude this by
inserting the explicit expression for $p_d$ and using the
Euler-Lagrange equation of motion for $\lb$.

Finally we observe, that a Killing vector $\xi^{\mu}$ of the
$d$-dimensional subspace can be lifted to a Killing vector of the
full $(d+1)$-dimensional space-time by taking \be \xi_A = (0, ...,
0, \bg) ~~~ \Leftrightarrow ~~~ \xi^A = (\xi^{\mu}, 1).
\label{3.1} \ee In this model, as all affine connection components
$\Gam_{AB}^{\;\;\;\;d} = 0$ and $\bg$ is a constant, it follows
immediately that eq.\ (\ref{1.7}) is satisfied.

We thus arrive at the following significant conclusion: The
generalized particle dynamics in $d$-dimensional space-time is
equivalent to a special class of geodesic motions in a
$(d+1)$-dimensional space-time with metric (\ref{2.1}),
which is an outcome of Kaluza-Klein decomposition, with
$n_{\mu} = \xi_{\mu}$, and characterized by $p_d = 0$.

\section{\bf Modeling Dark Energy}

Having convinced that the choice of radial function (\ref{i.1})
in the framework of generalized particle dynamics (\ref{action})
can be obtained through Kaluza-Klein decomposition technique, we now
turn to its cosmological implications. In order to obtain
cosmologically relevant conclusions, both from
theoretical and observational ground, we need to formulate
a cosmological model and estimate physically observable
quantities in $(3+1)$-dimensions. This is done by embedding a
$(3+1)$-dimensional FLRW space-time
into the $(4+1)$-dimensional effective metric (\ref{i.1})
and find out its consequences (for details of the formalism, see (\cite{brde, bwg})).
Thus, as the observable universe is $(3+1)$-dimensional,
we ultimately land up with a $(3+1)$-dimensional cosmological
scenario embedded in a $(4+1)$-dimensional background.
That is why we chose the background spacetime to be
$(4+1)$-dimensional as our starting point.

The $(3+1)$-dimensional FLRW metric in terms of the coordinates
$(T, \sg, \thg, \vf)$ is
\be
ds_{3+1}^2 = - dT^2 + a(T)^2 \left[ \frac{d\sg^2}{1 - k\sg^2}
 + \sg^2 (d\thg^2 + \sin^2 \thg\, d\vf^2) \right],
\label{j.5} \ee
Friedmann equation in a cosmological model with cosmological
constant $\Lambda$ is given by:
\be \frac{\dot{a}^2}{a^2} = -\frac{k}{a^2} + \frac{\Lambda}{3} +
\frac{8 \pi G_{3+1} \rho}{3}. \label{frw} \ee
For embedding in the brane world scenario, we exploit the well-known embedding
mechanism with the Gauss-Codazzi junction conditions and $Z_2$
symmetry \cite{gpdsch, bwg} and incorporate the usual practice of
identifying $r(T)$ with $a(T)$. Some comments regarding the application
of Gauss-Codazzi conditions in the present context is discussed in Section V.
Using these junction conditions and $Z_2$ symmetry, we then arrive at the
modified Friedmann equation for our model,
with spatially flat universe (k=0), consistent with
energy-conservation of ordinary matter on the brane, as
\be \frac{\dot{a}^2}{a^2} = -\frac{F(a)}{a^2} + \left(\frac{8 \pi
G_{4+1} \rho_{3+1}}{3}\right)^2. \label{frw1}\ee
Using the decomposition $\rho_{3+1} = \rho_0 + \rho$, where $\rho$
is the ordinary matter density on the brane and $\rho_0$ is the
brane tension, the modified Friedmann equation (\ref{frw1}) now becomes:
\be \frac{\dot{a}^2}{a^2} = -\frac{F(a)}{a^2} + \left(\frac
{8 \pi G_{3+1}}{3}\right)\left[\frac{1}{2} \rho_0 + \rho +
\frac{1}{2} \frac{\rho^2}{\rho_0}\right], \label{frw2} \ee
where $$G_{3+1} = \left(\frac{16 \pi G_{4+1}^2 \rho_0}{3}\right).$$

For confinement of matter on the brane (so that the brane matter
does not escape into the bulk freely), the brane tension $\rho_0$
is considered to be much larger than $\rho$, i.e. $\rho_0 \gg \rho$.
Hence the $\rho^2$ term in the Friedmann equation (\ref{frw2}) is
suppressed. Then the final form of the modified Friedmann equation (\ref{frw2}) is:
\be \frac{\dot{a}^2}{a^2} = -\frac{F(a)}{a^2} + \frac{4 \pi G_{3+1}
\rho_0}{3} + \frac{8 \pi G_{3+1} \rho}{3}. \label{mfrw} \ee This
modified Friedmann equation (\ref{mfrw}) is then compared with the
normal Friedmann equation (\ref{frw}) for further study in cosmology.\\
For our model, this $F(a)$ in (\ref{mfrw}) is the radial function
(\ref{i.2}) of the induced Schwarzschild-AdS metric (\ref{i.1})
and hence in our case, the modified Friedmann equations turns out to be
\bea
\left(\frac{\dot{a}}{a}\right)^2 = \frac{8\pi
G_{3+1}}{3}\rho+\frac{2M}{a^4}+ \left[ \alpha-\Lambda_5
+\frac{l^2(2M/a^2 -\Lambda_5a^2)}{\beta(a^2+l^2)}\right]
\label{fried1} \\
\frac{\ddot{a}}{a} = - \frac{4\pi
G_{3+1}}{3}(\rho+3p)-\frac{2M}{a^4}+ \left[ \alpha-\Lambda_5
-\frac{l^2(\Lambda_5  a^4+2M+2\Lambda_5 l^2 a^2)}{\beta(a^2+l^2)^2}\right]
\label{fried2}
\eea
where $\dot a = da/dT$ and $\alpha=\left(8\pi G_{4+1}\rho_0/3 \right)^2$.
Identifying this equation (\ref{fried1})
with (\ref{mfrw}), it becomes clear that except the first term $\frac{8\pi
G_{3+1}}{3}\rho$, which is also present in normal Friedmann equation (\ref{frw}),
all other terms in the r.h.s. of (\ref{fried1}) are originated either from
$F(a)$ or the brane tension $\rho_0$ of the modified Friedmann equation
(\ref{mfrw}). So these terms are geometrical ((4+1)-dimensional geometry) in nature.
Thus, for our cosmological model, only the knowledge about the bulk geometry
is important. Once somehow the bulk geometry is known (as is the case in our
model), the bulk source term ((4+1)-dimensional energy-momentum tensor)
becomes irrelevant for the field equations on the brane (Einstein
equation for brane) since there is no energy-momentum (matter)
exchange between the brane and the bulk.\\
The terms containing $M$ ($\propto a^{-4}$) in (\ref{fried1}, \ref{fried2}) contribute to
the radiation energy density of the universe (constrained by Nucleosynthesis
data to contribute to $\leq .03 \%$ of total
radiation energy density \cite{bwg}). We express the Hubble parameter
$H = \dot{a}/a$ in terms of the redshift $z$ (where $1+z =a_0/a= 1/a$) and
neglect contributions from any cosmic constituent which redshifts away at
the rate of radiation or faster (order of $(1+z)^4$ or higher) for a
late time universe, to express the Friedmann equation (\ref{fried1}) in a convenient form:
\begin{equation}
H^2=H_0^2\left[\Omega_X\left(1+b(1+z)^2\right)+\Omega_M(1+z)^3\right]
\label{H}
\end{equation}
where $\Omega_M(1+z)^3=8\pi G_{3+1} \rho/3H_0^2$ and $\Omega_X = (\alpha-\Lambda_5 \mu^2)/H_0^2$
denote respectively the density parameters for the matter sector and
for the additional terms coming from our modified gravity theory,
with the dimensionless parameter $b = \Lambda_5 \beta (\mu^2 -1)^2/ (\alpha-\Lambda_5 \mu^2)$.
Observations fix $H_0 = 70.5 \pm 1.3 ~{\rm km/s/Mpc}$ from the WMAP5 data
\cite{wmap5} and $H_0=74.2 \pm  3.6 ~{\rm km/s/Mpc}$ from the SHOES Team
data \cite{shoes}. Eq. (\ref{H}) is the major result of our letter, the
implications of which we analyze below.

Physical implications of the parameters are as follows:
$\Omega_M$, the sum-total of the density-fraction for luminous and
dark matter contributes $0.28 \pm 0.08$ of total cosmic density, as
fixed independently by the CMB \cite{hstcmb} and large scale structure
data \cite{lsst}. Hence, for a valid dark energy model, $\Omega_X$
accounts for the DE density. However, observations indicate a universe
close to the $\Lambda$CDM model. This forces $b$ to be small and
$\beta$ large ($\mu$ contains $\beta$ in the denominator).
Observationally, these values will be restricted by $\chi^2$
fitting, which we discuss later.

A crucial part from observational ground is to develop and estimate the
observable parameters to show that we have a late accelerating
universe where $\Omega_X$  accounts for the dark energy density.
We show that, consistent with observations, our candidate dark
energy EOS $(w_X)$ indeed crosses the phantom divider.\\

{\bf{Luminosity-redshift relation:}} This  determines dark energy
density $\Omega_{X}$ from observations, and, for our model
(\ref{H}), is given by
 \be d_L(z) = (1+z)
\int_{0}^{z} \frac{dz'}{H(z')} =  \frac{(1+z)}{H_0} \int_{0}^{z}
\frac{dz'}{\left[\Omega_X\left(1+b(1+z')^2\right)+\Omega_M(1+z')^3\right]^{1/2}}
\label{Lz}. \ee

\begin{figure}[htb]
{\centerline{\includegraphics[width=9cm, height=6cm] {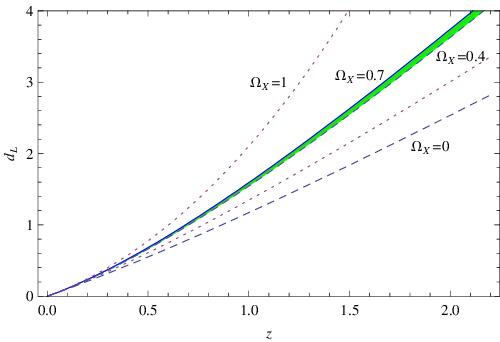}}}
\caption{Variation of the luminosity distance with redshift for
different $\Omega_{X}$}
\label{figLz}
\end{figure}

In Figure \ref{figLz} the variation of $d_L$ with redshift in units of
$H_0^{-1}$ for different $\Omega_{X}$ (with $\Omega_M +\Omega_{X}
=1$) is shown using numerical integration. The plots show that a
universe with $\Omega_{X} \sim 0.7,~ \Omega_M \sim 0.3$ is favored
in this model, and confirms that $\Omega_{X}$ accounts for the
dark energy density. The allowed region (dark shade) gives the
bound for $b$ as $-0.07 \leq b <0$.  Throughout the rest of the
paper, we take a representative small negative value for $b =
-0.05$. Once the luminosity distance is estimated, the apparent
magnitude of the Supernovae can be calculated from the Hubble
constant-free distance modulus \cite{de} $\mu(z) = 5 \log_{10}
\left[d_L(z)/{\rm Mpc} \right] +25$. With $d_L$ in Eq.\
(\ref{Lz}), we have checked that the plot for this quantity too
matches observations.\\

{\bf{Age of the universe:}}  It has the currently accepted value
of $13.7 \pm 0.02$ Gyr \cite{hstcmb} and (up to a certain redshift
$z$), is expressed as
\be t(z) = \int_{z}^{\infty}
\frac{dz'}{(1+z')H(z')} = \frac{1}{H_0} \int_{z}^{\infty}
\frac{dz'}{(1+z')\left[\Omega_X\left(1+b(1+z')^2\right)+\Omega_M(1+z')^3\right]^{1/2}}.
\ee

\begin{figure}[htb]
{\centerline{\includegraphics[width=9cm, height=6cm] {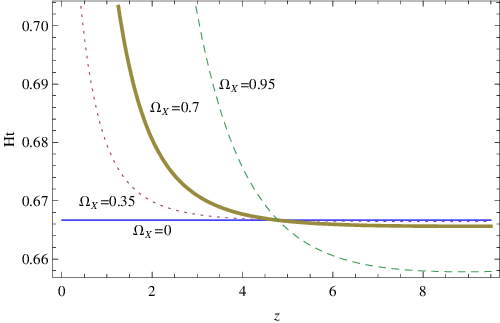}}}
\caption{Variation of H(z)t(z) with z in units of $H_0^{-1}$}
\label{figHtz}
\end{figure}
In Figure \ref{figHtz} we numerically plot $H(z)t(z)$ against redshift,
showing that $\Omega_{X} \sim 0.7$ represents the most acceptable
behavior.\\

{\bf{Deceleration parameter:}} This  will explicitly show the late
accelerating behavior. From (\ref{H}) we have
\begin{equation}
q(z)=\frac{-\ddot{a}/a}{\dot{a}^2/a^2}=\frac{H'(z)}{H(z)}(1+z)-1
=\frac{\Omega_M(1+z)^3-2\Omega_X}{2(\Omega_X(1+b(1+z)^2)+\Omega_M(1+z)^3)}.
\end{equation}

\begin{figure}[htb]
{\centerline{\includegraphics[width=9cm, height=6cm] {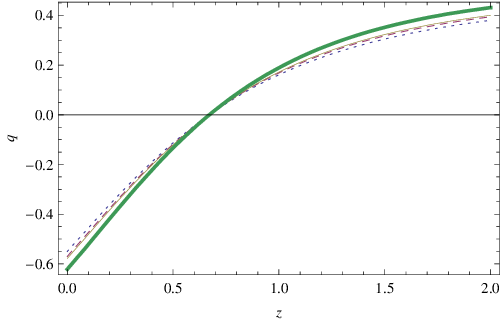}}}
\caption{Variation of the deceleration parameter $q$ with
redshift $z$ for different values of $b$. The dotted, dashed, thin and thick
lines correspond respectively to the values $b = 0, -.05, -.07, -.165$.}
\label{figqz}
\end{figure}

In Figure \ref{figqz}, the deceleration parameter has been plotted against
redshift for $\Omega_{X} \sim 0.7, \Omega_M \sim 0.3$ with different values
of $b$. The plots confirm that our model indeed results in an early decelerating
and late accelerating universe. Moreover, onset of the recent
accelerating phase, when the universe was $\sim 60 \%$ of its
present size ($z = 0.6$), is also confirmed by our model.\\

{\bf{Equation of state (EOS):}} The effective EOS of DE  in our
model is \be w_X(z)= \frac{2q(z)-1}{3[1-\Omega_M(z)]}
=-1+\frac{2b(1+z)^2}{3}. \ee The expression has been obtained by a
binomial expansion considering small $b$ and dropping terms of
order $(1+z)^4$ or higher, as before. Obviously, since $b$ is
negative and non-zero, the effective EOS of dark energy candidate
satisfies $w_X<-1$. So it shows a phantom like behavior that is
as claimed by SN1a Gold data set \cite{pdlobs}.

\begin{figure}[htb]
{\centerline{\includegraphics[width=9cm, height=6cm] {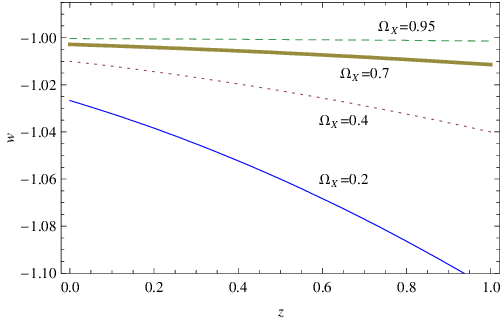}}}
\caption{Variation of the effective EOS of dark energy with
redshift}
\label{figwz}
\end{figure}

In Figure \ref{figwz}, the effective EOS parameter of dark energy with
redshift is shown for different values of $\Omega_X$. The latest
WMAP5 data constrains the lower bound of the dark energy EOS today
to be $-1.11 < w_{DE}$ \cite{wmap5}. This sets the lower bound of
$b$ at  $(-.165 \leq b)$, which is way below its lower bound $(-.07
\leq b)$ as predicted before from the luminosity-redshift
relation. This model with $(-.07 \leq b < 0)$ will thus fit well
with a more precise bound for the EOS available in the future.\\

{\bf{Statefinder parameters $\{r,s\}$:}} The parameters \cite{statefinder}
\be
r=\frac{\dddot a/a}{(\dot a/a)^3}
=1+\left[\frac{H''}{H}+\left(\frac{H'}{H}\right)^2\right](1+z)^2-2\frac{H'}{H}(1+z)
~ ;~ s=\frac{2}{3}\frac{r-1}{2q-1} \ee can distinguish dynamical
models from $\Lambda$CDM. In our model,  $\{r,s\}$ pair is \be
r=1-\frac{b \Omega_X(1+z)^2}{\Omega_X(1+b(1+z)^2)+\Omega_M(1+z)^3}
~ ;~
 s=\frac{2}{3} \frac{b(1+z)^2}{\left[ 3+b(1+z)^2\right] }
\ee

\begin{figure}[htb]
{\centerline{\includegraphics[width=9cm, height=6cm] {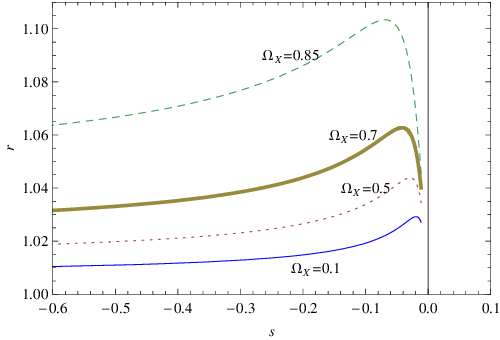}}}
\caption{Variation of r with s}
\label{figrs}
\end{figure}

In figure \ref{figrs} we plot $r$ versus $s$. The one with $\Omega_X = 0.7$
is the most favored plot in our model. Hence, in so far as $b$ is
non-trivial (as in the present case), the $\{r,s\}$ pair can
distinguish our alternative gravity model from $\Lambda$CDM (for
which $r = 1, s = 0$).

\section{Summary and outlook}

Let us summarize. We have constructed a specific modified gravity
structure, induced by a higher dimensional non-minimal particle
dynamics framework. This particular generalized relativistic
particle model was introduced by us in \cite{gpdsch}. Subsequently
we consider brane models embedded in this modified (4+1)-dimensional
AdS-Schwarzschild spacetime. Note that we are dealing
with an effective theory of gravity and its applications to cosmology,
since derivation of our induced (4+1)-dimensional metric,
generated by an yet unknown source term, remain an open issue.
Indeed, to project the DE feature, it is bound to be of a novel nature.
Furthermore, this issue is closely related to the use of
Gauss-Codazzi equations in Section IV, which requires the
metric to be a solution of Einstein's equation. Our analysis in section
III, in the Kaluza-Klein framework indicates that it is reasonable to expect
our effective metric to be compatible with Einstein's equation with a
suitable source term.\\
The most promising feature of our work lies in the fact that
the induced particle dynamics can
simulate cosmological evolution with a late time accelerating universe.
Not only that; in our formulation, it is also possible to have an effective phantom
dark energy model {\it{without invoking the phantom}}. This is
(quantitatively) manifested in the crossover of the phantom
barrier. Our model is qualitatively distinct, but not
quantitatively far off, from the $\Lambda$CDM model during recent
times. Hence all the positive features of $\Lambda$CDM along with
a phantom behavior (without the problems related to the negative
kinetic term) can be accommodated in our model.\\
Several aspects of the proposed framework can be investigated further. The parameters
used in the model can be constrained observationally by using a
maximum likelihood method involving the minimization of the
function $\chi^2 = \sum_{1= 1}^{N} \left[d_L(z)_{\rm obs} -
d_L(z)_{\rm th}\right]^2/\sigma_i^2$, where $d_L(z)_{\rm th}$
contains the parameters used in a specific theory, $N$ is the
number of Supernovae (taken as 157 for the most reliable Gold
dataset) and $\sigma_i$ are the $1\sigma$ errors from the
observational method used \cite{de}. Observationally, this is the
most accurate probe of $\Omega_{\rm DE}$; it will further
constrain the parameters used in our model. Further, the variable
EOS may be reflected in the Integrated Sachs-Wolfe (ISW) effect,
which will serve as another test for the model. Studying features
related to perturbations in this cosmological framework is another
open issue.

\end{document}